\RequirePackage[2020-02-02]{latexrelease}
\documentclass[onecolumn,amsmath,amssymb,10pt,aps]{revtex4}
  
\usepackage[utf8]{inputenc}
\usepackage[T1]{fontenc}

\usepackage{amsmath}
\usepackage{amsfonts}
\usepackage{graphicx}
\usepackage{yfonts}
\usepackage{color}
\usepackage[normalem]{ulem}
\usepackage{amsthm}
\usepackage{bm}
\usepackage{bbm}
\usepackage{mathtools}
\usepackage{array}
\usepackage{placeins}
\usepackage{enumitem}

\newcommand{\der}{\,\mathrm{d}}

\def\<{\langle}
\def\>{\rangle}
\newcommand{\tr}{\mathrm{Tr}}
\newcommand{\Tr}{\mathrm{Tr}}
\def\oper{{\mathchoice{\rm 1\mskip-4mu l}{\rm 1\mskip-4mu l}
{\rm 1\mskip-4.5mu l}{\rm 1\mskip-5mu l}}}
\DeclareMathAlphabet\mathbfcal{OMS}{cmsy}{b}{n}

\mathchardef\mhyphen="2D 

\newtheorem{Proposition}{Proposition}
\newtheorem{Example}{Example}

\begin{document}

\title{Phase-covariant mixtures of non-unital qubit maps}

\author{Katarzyna Siudzi\'{n}ska}
\affiliation{Institute of Physics, Faculty of Physics, Astronomy and Informatics, Nicolaus Copernicus University in Toru\'{n}, ul.~Grudzi\k{a}dzka 5, 87--100 Toru\'{n}, Poland}

\begin{abstract}
We analyze convex combinations of non-unital qubit maps that are phase-covariant. In particular, we consider the behavior of maps that combine amplitude damping, inverse amplitude damping, and pure dephasing. We show that mixing non-unital channels can result in restoring the unitality, whereas mixing commutative maps can lead to non-commutativity. For the convex combinations of Markovian semigroups, we prove that classical uncertainties cannot break quantum Markovianity. Moreover, contrary to the Pauli channel case, the semigroup can be recovered only by mixing two other semigroups.
\end{abstract}

\flushbottom

\maketitle

\thispagestyle{empty}

\section{Introduction}

In the theory of open quantum systems, the evolution of a physical system is described using dynamical maps $\Lambda(t)$. By definition, $\Lambda(t)$ are time-parameterized families of quantum channels (completely positive, trace-preserving maps) satisfying the initial condition $\Lambda(0)=\oper$ \cite{Breuer}. They transform any input state $\rho$ into an output state $\rho(t)=\Lambda(t)[\rho]$ at a time $t>0$.
By assuming weak coupling between the system and the environment and separation of characteristic time scales, one can use the Born-Markov approximation to derive the Markovian master equation \cite{BreuerPetr}
\begin{equation}
\dot{\Lambda}(t)=\mathcal{L}\Lambda(t).
\end{equation}
The most general generator $\mathcal{L}$ has the Gorini-Kossakowski-Sudarshan-Lindblad (GKSL) form \cite{GKS,L}
\begin{equation}
\mathcal{L}[\rho]=-i[H,\rho]+\sum_\alpha\gamma_\alpha
\left\{V_\alpha\rho V_\alpha^\dagger-\frac 12 \left[V_\alpha^\dagger V_\alpha,\rho\right]_+\right\},
\end{equation}
where $H$ is the effective Hamiltonian, $V_\alpha$ denote the noise operators, and the decoherence rates $\gamma_\alpha$ are positive numbers. To go beyond the Markovian master equation, memory effects caused by non-trivial interactions with the environment have to be included \cite{RHP,BLP,Vega}. One way to accomplish this is by introducing a time-local generator $\mathcal{L}(t)$, which has the GKSL form but with time-dependent $H(t)$, $V_\alpha(t)$, and $\gamma_\alpha(t)$.

One definition of quantum Markovianity is related to divisibility of dynamical maps \cite{RHP,Wolf}. Recall that $\Lambda(t)$ is divisible if for any $t\geq s\geq 0$ there exists a map $V(t,s)$ (propagator) such that
\begin{equation}\label{CP-div}
\Lambda(t)=V(t,s)\Lambda(s).
\end{equation}
If $V(t,s)$ is always positive, then the corresponding $\Lambda(t)$ is a P-divisible map. Analogically, CP-divisible $\Lambda(t)$ has a completely positive propagator and describes a Markovian evolution \cite{RHP,Sabrina}. Moreover, if $\Lambda(t)$ is a solution of the master equation with a time-local generator $\mathcal{L}(t)$, then $\Lambda(t)$ is CP-divisible if and only if $\gamma_\alpha(t)\geq 0$. Otherwise, the evolution is non-Markovian, which means that the coupling between the system and its environment is so strong that the effects of memory are no longer negligible \cite{Bernardes,JinGiovannetti,WuHou,Walborn}. Quantum evolution with memory effects is a modern research area that has experienced rapid development in recent years \cite{BLPV,Alonso}. It finds a wider ange of applications in quantum information processing,
quantum communication \cite{BreuerPetr,RivasHuelga,Nielsen}.

To simplify the evolution equations, one introduces symmetries to the dyncamical maps. Consider a finite group $G$ along with its two unitary representations $U_k$ on the Hilbert spaces $\mathcal{H}_k$, $k=1,2$. By definition, a linear map $\Lambda:\mathcal{B}(\mathcal{H}_1)\to \mathcal{B}(\mathcal{H}_2)$ is (unitarily) covariant with respect to $U_1$ and $U_2$ if
\begin{equation}\label{cov_def}
\Lambda\big[U_1(g)X U_1^\dagger(g)\big] = U_2(g)\Lambda[X]U_2^\dagger(g)
\end{equation}
for all operators $X \in \mathcal{B}(\mathcal{H}_1)$ and group elements $g \in G$. By extension, if such $\Lambda$ is a completely positive, trace-preserving map, then it is called the {\it covariant quantum channel}. The notion of unitarily covariant maps was first mentioned by Scutaru, who proved the Stinespring-type theorem for completely positive covariant maps \cite{Scutaru}. Covariant quantum channels were analyzed by Holevo along with covariant Markovian generators \cite{Holevo1993,CQME}. There are two channels covariant with respect to any unitary representations: depolarizing channels \cite{Keyl} and transpose depolarizing channels \cite{additiv,DHS}. Examples of quantum channels covariant only with respect to a selected unitary basis include the Pauli channels, the Weyl channels (also called {\it Weyl-covariant}) \cite{CQME,Amosov,Holevo2005} and generalized Pauli channels \cite{ICQC}.

Another special case of covariant channels are phase-covariant qubit maps, which are covariant with respect to $U(\phi)=\exp(-i\sigma_3\phi)$ for all real parameters $\phi$. Such channels describe any evolution that arises from a combination of pure dephasing with energy absorption and emission \cite{phase-cov-PRL,phase-cov}. Initially, the master equation for phase-covariant dynamical maps was introduced phenomenologically to characterize thermalization and dephasing processes beyond the Markovian approximation \cite{PC1}. An explicit microscopic derivation was provided using a weakly-coupled spin-boson model under the secular approximation \cite{PC3}. Futher studies showed a connection between the population monotonicity, coherence monotonicity, and Markovianity \cite{PC2}. Non-Markovian evolution of phase-covariant channels was also analyzed in ref. \cite{phase-cov}, where the authors presented examples of eternally non-Markovian evolution for non-unital, non-commutative dynamical maps.

Recently, convex combinations of legitimate dynamical maps have been given a significant attention. These are special classes of quantum maps that arise from classical mixtures of quantum channels. Many scenarios have been considered so far, such as mixing quantum maps that are Markovian semigroups \cite{ENM,Nina,mub_final,ICQC,Jagadish2}, CP-divisible \cite{CCMS,Jagadish3}, CP-indivisible \cite{CCnM}, or even non-invertible \cite{CCMK,Noninvertibility,CC_div}. However, all this was done only for the Pauli and generalized Pauli channels, which are both unital (preserve the maximally mixed state).

In this paper, we go beyond mixtures of unital maps and analyze convex combinations of phase-covariant qubit maps. Section 2 presents a quick introduction to phase-covariant channels, their complete positivity conditions, and the corresponding time-local generators. In Section 3, we consider mixtures of Markovian semigroups, proving that non-unital maps can give rise to the maps that are unital but not vice versa. Next, we analyze convex combinations of both invertible and non-invertible dynamical maps. Here, we prove that non-commutative maps can be mixed into commutative ones. Comparisons with convex combinations of Pauli channels are made. It turns out that mixtures of phase-covariant maps manifest significantly different behaviors.

\section{Phase-covariant qubit channels}

The most general form of the phase-covariant qubit channel reads
\cite{phase-cov-PRL,phase-cov}
\begin{equation}
\Lambda[X]=\frac 12 \left[(\mathbb{I}+\lambda_\ast\sigma_3)\tr X+\lambda_1\sigma_1
\tr(\sigma_1X)+\lambda_1\sigma_2\tr(\sigma_2X)+\lambda_3\sigma_3\tr(\sigma_3X)
\right],
\end{equation}
where $\sigma_\alpha$ are the Pauli matrices. Moreover, $\lambda_1$ and $\lambda_3$ are two of its eigenvalues ($\lambda_1$ is two-times degenerate) to the eigenvectors determined as in the eigenvalue equations
\begin{equation}
\Lambda[\sigma_1]=\lambda_1\sigma_1,\qquad 
\Lambda[\sigma_2]=\lambda_1\sigma_2,\qquad 
\Lambda[\sigma_3]=\lambda_3\sigma_3. 
\end{equation}
The last eigenvalue equation
\begin{equation}
\Lambda[\rho_\ast]=\rho_\ast
\end{equation}
determines the state $\rho_\ast$ preserved by $\Lambda$, which is given by the formula
\begin{equation}
\rho_\ast=\frac 12 \left[\mathbb{I}+\frac{\lambda_\ast}{1-\lambda_3}\sigma_3\right],
\end{equation}
and therefore it depends on the parameter $\lambda_\ast$ and the channel eigenvalue $\lambda_3$. Whenever $\lambda_\ast$ is non-zero, $\Lambda$ is a non-unital map ($\Lambda[\mathbb{I}]\neq\mathbb{I}$). Note that $\lambda_1$, $\lambda_3$, and $\lambda_\ast$ are all real due to the Hermiticity of $\sigma_\alpha$. The complete positivity conditions for $\Lambda$ read
\begin{equation}
|\lambda_3|+|\lambda_\ast|\leq 1,\qquad 4\lambda_1^2+\lambda_\ast^2\leq(1+\lambda_3)^2.
\end{equation}
Finally, observe that two phase-covariant channels $\Lambda_1$, $\Lambda_2$ are not commutative in general; that is, $\Lambda_1\Lambda_2\neq\Lambda_2\Lambda_1$.
This property could not be observed for unital qubit (Pauli) channels.

Assume that the phase-covariant channel is a solution of a master equation $\dot{\Lambda}(t)=\mathcal{L}(t)\Lambda(t)$, $\Lambda(0)=\oper$, with the time-local generator, whose most general form is
\begin{equation}\label{generator}
\mathcal{L}(t)=\gamma_+(t)\mathcal{L}_++\gamma_-(t)\mathcal{L}_-+\gamma_3(t)\mathcal{L}_3,
\end{equation}
where $\gamma_\pm(t)$ and $\gamma_3(t)$ are the decoherence rates and
\begin{equation}
\mathcal{L}_{\pm}[X]=\sigma_\pm X\sigma_\mp -\frac 12 \{\sigma_\mp\sigma_\pm,X\},
\qquad \mathcal{L}_3[X]=\frac 14(\sigma_3X\sigma_3-X).
\end{equation}
This evolution includes several special cases, such as amplitude damping ($\gamma_1(t)=\gamma_3(t)=0$), generalized amplitude damping ($\gamma_3(t)=0$), and pure dephasing ($\gamma_1(t)=\gamma_2(t)=0$) \cite{Nielsen}. The relation between the decoherence rates and the eigenvalues of the corresponding dynamical map can be recovered from the eigenvalue equations for the generator
\begin{equation}\label{gen}
\begin{split}
\mathcal{L}(t)[\sigma_1]=&-\frac 12 (\gamma_++\gamma_-+\gamma_3)\sigma_1,\qquad
\mathcal{L}(t)[\sigma_3]=-(\gamma_++\gamma_-)\sigma_3,\\
\mathcal{L}(t)[\sigma_2]=&-\frac 12 (\gamma_++\gamma_-+\gamma_3)\sigma_2,
\end{split}
\end{equation}
and one additional equation, $\mathcal{L}(t)[\rho_\ast]=\dot{\rho}_\ast$.
Hence, one arrives at
\begin{equation}
\lambda_1(t)=\exp\left\{-\frac 12 \Big[\Gamma_+(t)+\Gamma_-(t)+\Gamma_3(t)\Big]\right\},\qquad
\lambda_3(t)=\exp\Big[-\Gamma_+(t)-\Gamma_-(t)\Big],
\end{equation}
\begin{equation}
\lambda_\ast(t)=\exp\Big[-\Gamma_+(t)-\Gamma_-(t)\Big]\int_0^t
\Big[\gamma_+(\tau)-\gamma_-(\tau)\Big]\exp\Big[\Gamma_+(\tau)+\Gamma_-(\tau)\Big]
\der\tau,
\end{equation}
where $\Gamma_\mu(t)=\int_0^t\gamma_\mu(\tau)\der\tau$, $\mu=\pm,3$.
Observe that only $\lambda_\ast(t)$ as asymmetric with respect to the change of signs in $\gamma_\pm(t)$. The inverse relation reads
\begin{equation}
\gamma_{\pm}(t)=\frac{\lambda_3(t)}{2}\frac{\der}{\der t}
\left(\frac{1\pm\lambda_\ast(t)}{\lambda_3(t)}\right),\qquad
\gamma_3(t)=\frac{\der}{\der t}
\ln\frac{\lambda_3(t)}{[\lambda_1(t)]^2}.
\end{equation}
The evolution provided by $\mathcal{L}(t)$ from eq. (\ref{generator}) is Markovian if and only if $\gamma_\pm(t)\geq 0$ and $\gamma_3(t)\geq 0$ for all $t\geq 0$. The Markovian semigroup is reproduced by positive, time-independent rates, and its eigenvalues satisfy the following formulas \cite{phase-cov},
\begin{equation}\label{MSG}
\lambda_1(t)=\exp\left\{-\frac t2 \Big[\gamma_++\gamma_-+\gamma_3\Big]\right\},\qquad
\lambda_3(t)=\exp\Big[-(\gamma_++\gamma_-)t\Big],
\end{equation}
\begin{equation}
\lambda_\ast(t)=\frac{\gamma_+-\gamma_-}{\gamma_++\gamma_-}
\left[1-e^{-(\gamma_++\gamma_-)t}\right].
\end{equation}

\section{Mixtures of non-unital qubit channels}

A special class of physical channels is a classical mixture of legitimate dynamical maps $\Lambda_\alpha(t)$ with probabilities $x_\alpha$,
\begin{equation}
\Lambda(t)=\sum_{\alpha=1}^Nx_\alpha\Lambda_\alpha(t).
\end{equation}
By construction, $\Lambda(t)$ is a valid dynamical map. So far, in the literature, only convex combinations of unital maps have been analyzed. However, mixtures of non-unital maps allow us to observe certain behaviors that did not occur when mixing unital maps. First, a mixture of unital maps always remains unital; however, the converse is no longer true.

\begin{Proposition}
A mixture of non-unital quantum maps can result in a unital map.
\end{Proposition}

\begin{proof}
Consider a convex combination $\Lambda(t)$ of $N$ phase-covariant qubit channels $\Lambda_\alpha(t)$, where
\begin{equation}
\Lambda(t)=\sum_{\alpha=1}^Nx_\alpha\Lambda_\alpha(t).
\end{equation}
Denote the eigenvalues and the parameter responsible for non-unitality that characterize $\Lambda_\alpha(t)$ by $\lambda_k^{(\alpha)}(t)$, $k=1,3$, and $\lambda_\ast^{(\alpha)}(t)$, respectively. Then, the action of the mixture on the identity operator $\mathbb{I}$ is given by
\begin{equation}
\Lambda(t)[\mathbb{I}]=\mathbb{I}+\sum_{\alpha=1}^Nx_\alpha\lambda_\ast^{(\alpha)}(t)
\sigma_3.
\end{equation}
Therefore, $\Lambda(t)$ is unital as long as $\sum_{\alpha=1}^Nx_\alpha\lambda_\ast^{(\alpha)}(t)=0$ at any time $t\geq 0$.
\end{proof}

\begin{Example}
Let us take a mixture
\begin{equation}
\Lambda(t)=\frac 12 \left[\Lambda_1(t)+\Lambda_2(t)\right]
\end{equation}
of two phase-covariant qubit channels. We choose $\Lambda_1(t)$ and $\Lambda_2(t)$ in such a way that they share all the eigenvalues ($\lambda_k^{(1)}(t)=\lambda_k^{(2)}(t)$). Finally, we fix their last defining parameters, $\lambda_\ast^{(1)}(t)$ and $\lambda_\ast^{(2)}(t)$, so that they only differ in signs ($\lambda_\ast^{(1)}(t)=-\lambda_\ast^{(2)}(t)$). In this case, one has
\begin{equation}
\Lambda(t)[\mathbb{I}]=\frac 12 [(\mathbb{I}+\lambda_\ast^{(1)}(t)\sigma_3)+
(\mathbb{I}-\lambda_\ast^{(1)}(t)\sigma_3)]=\mathbb{I},
\end{equation}
which shows that $\Lambda(t)$ is indeed unital despite being a mixture of two non-unital maps.
\end{Example}

\subsection{Mixing Markovian semigroups}




Let us consider convex combinations of three Markovian semigroups
\begin{equation}\label{mix_MS}
\Lambda(t)=x_1e^{2w_1\mathcal{L}_+t}+x_2e^{2w_2\mathcal{L}_-t}
+x_3e^{2w_3\mathcal{L}_3t},
\end{equation}
where $w_\alpha\geq 0$. The corresponding eigenvalues read
\begin{equation}
\lambda_1(t)=x_1e^{-w_1t}+x_2e^{-w_2t}+x_3e^{-w_3t},\qquad
\lambda_3(t)=x_1e^{-2w_1t}+x_2e^{-2w_2t}+x_3,
\end{equation}
and
\begin{equation}
\lambda_\ast(t)=x_1\left(1-e^{-2w_1t}\right)-x_2\left(1-e^{-2w_2t}\right).
\end{equation}
Observe that the parameter $w_3$ determines only the eigenvalue $\lambda_1(t)$. Moreover, $\lambda_\alpha(t)$ do not depend on a single $x_\alpha$, which was the case for the Pauli channels. This complicates the formula for the time-local generator $\mathcal{L}(t)$ of the mixture, whose decoherence rates read as follows,
\begin{equation}\label{rates}
\begin{split}
\gamma_+(t)=&\frac{2x_1}{x_1e^{-2w_1t}+x_2e^{-2w_2t}+x_3}
\left\{w_1e^{-2w_1t}\left[1-x_2\left(1-e^{-2w_2t}\right)\right]
+x_2w_2e^{-2w_2t}\left(1-e^{-2w_1t}\right)\right\},\\
\gamma_-(t)=&\frac{2x_2}{x_1e^{-2w_1t}+x_2e^{-2w_2t}+x_3}
\left\{x_1w_1e^{-2w_1t}\left(1-e^{-2w_2t}\right)
+w_2e^{-2w_2t}\left[1-x_1\left(1-e^{-2w_1t}\right)\right]\right\},\\
\gamma_3(t)=&\frac{2\sum_{\mu=1}^3 x_\mu e^{-w_\mu t}
\left\{\sum_{\nu=1}^3x_\nu e^{-2w_\nu t}(w_\mu-w_\nu)
+x_3\left[w_\mu\left(1-e^{-2w_3t}\right)+w_3e^{-2w_3t}\right]\right\}}
{\sum_{\alpha,\beta=1}^3x_\alpha x_\beta e^{-(2w_\alpha+w_\beta)t}}.
\end{split}
\end{equation}

For the Pauli channels, the mixture of Markovian semigroups could lead to a Markovian or non-Markovian evolution, depending on the choice of the parameters \cite{CCMS}. For the phase-covariant channels, this in no longer the case.

\begin{Proposition}
All mixtures of Markovian semigroups given in eq. (\ref{mix_MS}) are CP-divisible.
\end{Proposition}

\begin{proof}
It is easy to see that $\gamma_\pm(t)\geq 0$ due to them being sums of positive terms. To prove that $\gamma_3(t)\geq 0$, it is enough to show that the nominator is also a sum of positive terms, as the denominator is always positive. The second term in the nominator is obviously greater than zero, so let us focus on the first term. It can be rewritten into
\begin{equation}
\sum_{\mu,\nu=1}^3 x_\mu x_\nu e^{-2(w_\nu+w_\mu) t}(w_\mu-w_\nu)
=\sum_{\alpha>\beta}x_\alpha x_\beta e^{-(w_\alpha+w_\beta)t} (w_\beta-w_\alpha)
\left(e^{-w_\alpha t}-e^{-w_\beta t}\right),
\end{equation}
which is indeed a sum of positive terms.
\end{proof}


\begin{Example}\label{ExMSG}
For the simple case with $w_\alpha=w$ and $x_3=0$, the decoherence rates from eq. (\ref{rates}) simplify to
\begin{equation}
\gamma_+(t)=2wx_1,\qquad\gamma_-(t)=2wx_2,\qquad\gamma_3(t)=0.
\end{equation}
Observe that the corresponding $\mathcal{L}(t)$ is the generator of the Markovian semigroup $\Lambda(t)$ (generalized amplitude damping channel). Therefore, contrary to the Pauli channel case \cite{CCMK,Noninvertibility}, it is possible to obtain the Markovian semigroup from a mixture of semigroups.
\end{Example}

\subsection{Beyond the semigroups}


In this section, we analyze mixtures of dynamical maps that are more general than Markovian semigroups. Namely, let us take
\begin{equation}\label{mix_MS_2}
\Lambda(t)=x_1\Lambda_+(t)+x_2\Lambda_-(t)+x_3\Lambda_3(t),
\end{equation}
where
\begin{equation}
\begin{split}
\Lambda_+(t)[X]&=\frac 12 \left\{\Big[\mathbb{I}+ (1-\eta_1^2(t))\sigma_3\Big]\Tr X+\eta_1(t)(\sigma_1\Tr\sigma_1X
+\sigma_2\Tr\sigma_2X)+\eta_1^2(t)\sigma_3\Tr\sigma_3X\right\},\\
\Lambda_-(t)[X]&=\frac 12 \left\{\Big[\mathbb{I}- (1-\eta_2^2(t))\sigma_3\Big]\Tr X+\eta_2(t)(\sigma_1\Tr\sigma_1X
+\sigma_2\Tr\sigma_2X)+\eta_2^2(t)\sigma_3\Tr\sigma_3X\right\},\\
\Lambda_3(t)[X]&=\frac 12 \left[\mathbb{I}\Tr X+\eta_3(t)(\sigma_1\Tr\sigma_1X
+\sigma_2\Tr\sigma_2X)+\sigma_3\Tr\sigma_3X\right].
\end{split}
\end{equation}
These maps satisfy the complete positivity conditions for $|\eta_k(t)|\leq 1$, $k=1,2,3$, and they describe Markovian semigroups when
\begin{equation}
\eta_k(t)=e^{-w_kt}.
\end{equation}
The eigenvalues of $\Lambda(t)$ are given by
\begin{equation}\label{rozne_eta}
\lambda_1(t)=\sum_{\alpha=1}^3x_\alpha\eta_\alpha(t),\qquad
\lambda_3(t)=x_1\eta_1^2(t)+x_2\eta_2^2(t)+x_3,
\end{equation}
and
\begin{equation}\label{rozne_eta2}
\lambda_\ast(t)=x_1[1-\eta_1^2(t)]-x_2[1-\eta_2^2(t)].
\end{equation}
Now, observe that it is admissible for $\eta_k(t)$ to reach zero at finite points in time. If this happens, then the corresponding dynamical map is non-invertible; i.e., the operator $\Lambda^{-1}(t)$ such that $\Lambda(t)\Lambda^{-1}(t)=\Lambda^{-1}(t)\Lambda(t)=\oper$ is not well-defined. From eq. (\ref{rozne_eta}), we see that mixtures of invertible maps always produce invertible $\Lambda(t)$. However, analogical statement does not hold for non-invertible maps.

\begin{Proposition}
A mixture $\Lambda(t)$ from eq. (\ref{mix_MS_2}) is an invertible dynamical map if and only if
\begin{equation}
\sum_{\alpha=1}^3x_\alpha\eta_\alpha(t)>0,\qquad x_1\eta_1^2(t)+x_2\eta_2^2(t)+x_3>0.
\end{equation}
\end{Proposition}

\begin{Example}
An example of non-invertible dynamical maps leading to an invertible mixture $\Lambda(t)$ follows for $x_1=x_2=x_3=1/3$ and
\begin{equation}
\eta_2(t)=\eta_3(t)=e^{-t},\qquad \eta_1(t)=e^{-t}\cos t.
\end{equation}
Note that, while $\Lambda_{\pm}(t)$ are always invertible, $\Lambda_3(t)$ is not due to the cosine function. In this case, the eigenvalues of $\Lambda(t)$, which read
\begin{equation}
\lambda_1(t)=\frac{e^{-t}}{3}(2+\cos t),\qquad
\lambda_3(t)=\frac 13 \left[1+e^{-2t}(1+\cos^2t)\right],\qquad
\lambda_\ast(t)=\frac{e^{-2t}}{3}\sin^2t,
\end{equation}
are always positive, and hence $\Lambda(t)$ is an invertible dynamical map.
\end{Example}

It has been shown that dynamical maps can be mixed to produce a semigroup \cite{CCMK}. In particular, for the Pauli channels, only a convex combination of three dephasing channels can result in a Markovian semigroup, of which at least two have to be non-invertible \cite{Noninvertibility}. A substantially different behavior can be observed for phase-covariant channels, which we discuss in more details below.

\begin{Proposition}
If $\Lambda(t)$ is a mixture of three dynamical maps, then it is not a Markovian semigroup.
\end{Proposition}

\begin{proof}
Following eq. (\ref{MSG}), $\Lambda(t)$ is a semigroup if and only if we choose $\eta_\alpha(t)$ in such a way that
\begin{equation}\label{prf}
\begin{split}
\eta_1(t)&=\sqrt{1-\frac{\gamma_+}{x_1(\gamma_++\gamma_-)}
\left(1-e^{-(\gamma_++\gamma_-)t}\right)},\\
\eta_2(t)&=\sqrt{1-\frac{\gamma_-}{x_2(\gamma_++\gamma_-)}
\left(1-e^{-(\gamma_++\gamma_-)t}\right)},\\
\eta_3(t)&=\frac{1}{x_3}\left[e^{-(\gamma_++\gamma_-+4\gamma_3)t}-x_1\eta_1(t)-x_2\eta_2(t)\right].
\end{split}
\end{equation}
The positivity of the terms under the square roots implies
\begin{equation}
x_1\geq\frac{\gamma_+}{\gamma_++\gamma_-},\qquad x_2\geq\frac{\gamma_-}{\gamma_++\gamma_-},
\end{equation}
and hence one has $x_1+x_2\geq 1$. This means that $x_3=0$, which was assumed to be non-zero. Hence, there are no valid solutions.
\end{proof}

\begin{Proposition}
The only mixture of two channels that produces a Markovian semigroup is the generalized amplitude damping channel given in Example \ref{ExMSG}. 
\end{Proposition}

\begin{proof}
First, assume that $x_3=0$. Again starting from eq. (\ref{MSG}), we see that $\Lambda(t)$ is a semigroup if and only if we choose $\eta_1(t)$ and $\eta_2(t)$ exactly like in eq. (\ref{prf}). Now, the formulas for $\lambda_1(t)$ in eqs. (\ref{MSG}) and (\ref{rozne_eta}) impose one additional condition for the decoherence rates,
\begin{equation}\label{one}
e^{-(\gamma_++\gamma_-+4\gamma_3)t}=x_1\eta_1(t)+x_2\eta_2(t),
\end{equation}
which has to hold for any $t\geq 0$. In the special case of $t\to\infty$, one has
\begin{equation}
0=x_1\sqrt{\frac{x_1\gamma_--x_2\gamma_+}{x_1(\gamma_++\gamma_-)}}
+x_2\sqrt{\frac{x_2\gamma_+-x_1\gamma_-}{x_2(\gamma_++\gamma_-)}},
\end{equation}
which gives $x_1\gamma_-=x_2\gamma_+$, or equivalently $\gamma_+=2wx_1$ and $\gamma_-=2wx_2$ for $w>0$. Substituting this into eq. (\ref{one}), we get $\gamma_3=0$. These are exactly the rates from Example \ref{ExMSG}.

Now, if $x_1=0$, then
\begin{equation}
\eta_2(t)=\sqrt{\frac{e^{-(\gamma_++\gamma_-)t}-x_3}{x_2}},
\end{equation}
which holds for every $t\geq 0$ only for $x_3=0$, where there is no mixing of maps. Analogical results follow for $x_2=0$.
\end{proof}

In the special case where $\eta_k(t)=\eta(t)$, $k=1,2,3$, the eigenvalues of $\Lambda(t)$ simplify to
\begin{equation}
\lambda_1(t)=\eta(t),\qquad
\lambda_3(t)=x_3+(1-x_3)\eta^2(t),\qquad
\lambda_\ast(t)=(x_1-x_2)[1-\eta^2(t)].
\end{equation}
Note that the singularity point of $\lambda_1(t)$ is the same as of $\eta(t)$, and $\lambda_3(t)$ is always non-singular. The corresponding decoherence rates  are always of the same sign as
\begin{equation}
\begin{split}
\gamma_{+}(t)=&\frac{-2x_1\dot{\eta}(t)\eta(t)}
{x_3+(1-x_3)\eta^2(t)},\\
\gamma_{-}(t)=&\frac{-2x_2\dot{\eta}(t)\eta(t)}
{x_3+(1-x_3)\eta^2(t)},\\
\gamma_3(t)=&-\frac{\dot{\eta}(t)}{\eta(t)}\frac{2x_3}{x_3+(1-x_3)\eta^2(t)}.
\end{split}
\end{equation}
Finally, observe that $\Lambda_{\pm}(t)$ and $\Lambda_3(t)$ are invertible if and only if $\eta(t)>0$. Moreover, their convex combination is always invertible. On the other hand, if $\Lambda_{\pm}(t)$ and $\Lambda_3(t)$ are non-invertible, then they always result in a non-invertible $\Lambda(t)$. This is another difference from the convex combinations of Pauli channels, where non-invertible maps could also produce invertible maps, and even semigroups \cite{CCMK}.

A dynamical map $\Lambda(t)$ is commutative if $\Lambda(t)\Lambda(s)=\Lambda(s)\Lambda(t)$. To obtain an equivalent condition in terms of its eigenvalues, it is enough to check the action on the identity.
This way, we arrive at
\begin{equation}
\lambda_\ast(t)[1-\lambda_3(s)]=\lambda_\ast(s)[1-\lambda_3(t)].
\end{equation}
For the mixtures $\Lambda(t)$ with the eigenvalues given by eq. (\ref{rozne_eta}), this condition reduces to
\begin{equation}
(1-\eta_1^2(t))(1-\eta_2^2(s))=(1-\eta_1^2(s))(1-\eta_2^2(t)).
\end{equation}
Therefore, $\Lambda(t)$ is commutative if and only if
\begin{equation}
\eta_2^2(t)=a\eta_1^2(t)+1-a
\end{equation}
with a constant $a\geq 0$. As $\Lambda_\pm(t)$ and $\Lambda_3(t)$ are all commutative, it is evident that a mixture of commutative dynamical maps can lead to a non-commutative map; e.g., $\Lambda(t)$ from Example 3.

\section{Conclusions}

We analyzed mixtures of non-unital maps on the example of phase-covariant qubit maps. In particular, we considered combinations of amplitude damping, inverse amplitude damping, and pure dephasing. It was proven that non-unital channels can be mixed into unital channels, as well as mixing commutative maps can results in the maps that are non-commutative. For the convex combinations of Markovian semigroups, we showed that all resulting maps are Markovian (CP-divisible). Interestingly, one can only recover the Markovian semigroup by mixing two semigroups for amplitude damping and inverse amplitude damping. This behavior differs from the Pauli channels case, where the semigroup followed only from non-invertible maps. It would be interesting to further explore mixtures of non-unital dynamical maps by considering more general channels. For qudit systems, convex combinations of quantum maps were analyzed only for the generalized Pauli channels. One could wonder whether there would be just as many differences between these maps and mixtures of non-unital qudit channels.

\section{Acknowledgements}

This paper was supported by the Foundation for Polish Science (FNP) and the Polish National Science Centre project No. 2018/30/A/ST2/00837.

\bibliography{C:/Users/cyndaquilka/OneDrive/Fizyka/bibliography}
\bibliographystyle{C:/Users/cyndaquilka/OneDrive/Fizyka/beztytulow2}

\end{document}